\documentclass[conference,compsoc]{IEEEtran}

\ifCLASSOPTIONcompsoc
  \usepackage[nocompress]{cite}
\else
  \usepackage{cite}
\fi

\ifCLASSINFOpdf
\else
\fi
\usepackage{glossaries}
\usepackage{hyperref}
\usepackage{multirow}
\usepackage{graphicx}
\usepackage{subcaption}
\usepackage{pdflscape}
\usepackage{algorithm}
\usepackage{algpseudocode}

\usepackage{rotating}

\algnewcommand\algorithmicforeach{\textbf{for each}}
\algnewcommand\algorithmicinput{\textbf{Input:}}
\algnewcommand\algorithmicoutput{\textbf{Output:}}
\algnewcommand\Input{\item[\algorithmicinput]}%
\algnewcommand\Output{\item[\algorithmicoutput]}%

\algdef{S}[FOR]{ForEach}[1]{\algorithmicforeach\ #1\ \algorithmicdo}

\begin{document}
\newacronym{WHO}{WHO}{World Health Organization}
\newacronym{GDP}{GDP}{Gross Domestic Product}
\newacronym{IOT}{IoT}{Internet of Things}
\newacronym{CPS}{CPS}{Cyber-Physical-Systems}
\newacronym{HITLCPS}{HITLCPS}{Human-in-the-Loop Cyber-Physical-Systems}
\newacronym{ISABELA}{ISABELA}{Iot Student Advisor and BEst Lifestyle Analyser}
\newacronym{GE}{GE}{Generic Enablers}
\newacronym{API}{API}{Application Program Interfaces}
\newacronym{BLE}{BLE}{Bluetooth Low Energy}
\newacronym{SDK}{SDK}{Software Development Kit}
\newacronym{SAM}{SAM}{Self-Assessment Manikin}
\newacronym{SVM}{SVM}{Support Vector Machine}
\newacronym{LOOCV}{LOOCV}{Leave-One-Out-Cross-Validation}
\newacronym{GPA}{GPA}{Grade Point Average}
\newacronym{ROC}{ROC}{Receiver Operating Characteristic}
\newacronym{AUC}{AUC}{Area Under Curve}
\newacronym{EMA}{EMA}{Ecological Momentary Assessment}
\newacronym{PCA}{PCA}{Principal Component Analysis}
%
\title{Automatically Assessing Students' Performance with Smartphone Data}


\author{\IEEEauthorblockN{J. Fernandes}
\IEEEauthorblockA{CISUC,\\
University of Coimbra,\\
Coimbra, Portugal\\jmfernandes@dei.uc.pt}
\and
\IEEEauthorblockN{J. Sá Silva}
\IEEEauthorblockA{INESC,\\
University of Coimbra,\\
Coimbra, Portugal\\sasilva@deec.uc.pt}
\and
\IEEEauthorblockN{A. Rodrigues}
\IEEEauthorblockA{Coimbra Business School\\Research Centre $|$ ISCAC,\\
Polytechnic of Coimbra\\
and University of Coimbra,\\
Coimbra, Portugal\\andre@iscac.pt}

\and
\IEEEauthorblockN{S. Sinche}
\IEEEauthorblockA{Department of Electronics\\Telecommunications and Networks, \\
Escuela Politecnica Nacional,\\
Quito, Ecuador\\soraya.sinche@epn.edu.ec}
\and

\IEEEauthorblockN{F. Boavida}
\IEEEauthorblockA{CISUC,\\
University of Coimbra,\\
Coimbra, Portugal\\boavida@dei.uc.pt}
}


%


\maketitle

\begin{abstract}
As the number of smart devices that surround us increases, so do the opportunities to create smart socially-aware systems. In this context, mobile devices can be used to collect data about students and to better understand how their day-to-day routines can influence their academic performance. Moreover, the Covid-19 pandemic led to new challenges and difficulties, also for students, with considerable impact on their lifestyle. In this paper we present a dataset collected using a smartphone application (ISABELA), which include passive data (e.g., activity and location) as well as self-reported data from questionnaires. We present several tests with different machine learning models, in order to classify students' performance. These tests were carried out using different time windows, showing that weekly time windows lead to better prediction and classification results than monthly time windows. Furthermore, it is shown that the created models can predict student performance even  with data collected from different contexts, namely before and during the Covid-19 pandemic. SVMs, XGBoost and AdaBoost-SAMME with Random Forest were found to be the best algorithms, showing an accuracy greater than $78\%$. Additionally, we propose a pipeline that uses a decision level median voting algorithm to further improve the models' performance,  by using historic data from the students to further improve the prediction. Using this pipeline, it is possible to further increase the performance of the models, with some of them obtaining an accuracy greater than 90\%.

\end{abstract}


%
\IEEEpeerreviewmaketitle

\section{Introduction}
\label{sec:introduction}

\gls{IOT} devices grant us the ability to create systems capable of interfacing with both the physical world and humans. Recent statistics point out that there are roughly 12 billion \gls{IOT} devices worldwide, with the prospect to double this number in the next few years \cite{IoTconn92:online}. These smart devices are capable of incorporating sensors and being connected to the Internet or other similar devices, which allows them to sense the environment and share the obtained data. Furthermore, these devices have enough processing power for onboard data processing or even for creating \gls{CPS} that can control physical phenomena. 

These characteristics, make the use of \gls{IOT} devices ideal for human-centred applications and systems in which humans are the main component of control loops, thus leading to \gls{HITLCPS} \cite{nunes2015survey}. In general, \gls{HITLCPS} systems are able to gather and process data pertaining to human actions, and infer intents and mental states (e.g., emotions, feeling and wants). This opens the way to the creation of intelligent and adaptable advice systems that can assist humans in their day-to-day lives. 

One possible \gls{HITLCPS} use case is that of a system for assessing students' academic performance and provide guidance. Students' performance can be affected by a variety of factors, from social-economic factors to psychological or even environmental factors \cite{hijazi2006factors}. As such, it is hard to predict which students will have lower or higher performance. Additionally, novel educational mechanism which relying on the use of online tools and materials have been shown to improve students' performance \cite{strelan2020flipped}. However, since a growing percentage of the learning process happens outside of the classroom, evaluating the students' progress and needs is getting harder for teachers. In this respect, any system that can identify and/or predict poor performance can be invaluable.

In the past three years, we have been affected by the Covid-19 pandemic, which affected more than 220 countries, with more than 530 million reported cases. Apart from the obvious human and economic cost, studies have also pointed out the psycho-social effects of the Covid-19 pandemic \cite{haleem2020effects}. The measures adopted in many countries to limit the spreading of the virus have led to significant changes in the daily routines of people. At some point in time, almost every affected country had to switch from in-person to online classes, which had substantial impact on the daily lives of students. Some studies analysed the effect of these changes. In Italy, a study was conducted with adolescents, to evaluate the effects of the pandemic in their lifestyle and emotional states\cite{buzzi2020psycho}. The results point to younger generations being more resilient to these changes and to uncertain situations. On the other hand, other studies have highlighted the negative effects of the ongoing pandemic and associated restrictions \cite{afonso2020impact}. Evaluating how these changes affect student performance is also an important challenge.

In this paper, we present an approach to assess and predict students' academic performance. Our proposal is based on automatically collected data using the \gls{ISABELA} system, a \gls{HITLCPS} system presented in \cite{fernandes2020isabela}. \gls{ISABELA} operates in an unobtrusive and passive manner, by using smartphones, smartwatches, and \gls{IOT} boxes as sensing mechanisms. Additionally, the system makes use of an interactive agent,  commonly known as chatbot, to close the loop and provide feedback to students. Based on the acquired data, 
we present a model that accepts students' aggregated data over 1-week periods and is able to predict their performance with $\pm80$\% accuracy. Furthermore, we present a pipeline that is able to use historic data to further improve the model performance.

The dataset used for the creation of the presented models is based on two distinct one-month trials executed on different years, namely, a first trial in May-June 2018, and a second trial in March-April 2021. Both trials were executed  at the Escuela Politecnica Nacional of Ecuador. During the 2021 trial, the Covid-19 pandemic was ongoing in Ecuador, and the classes were lectured remotely. As such, this dataset also offers an insight on how Covid-19-related changes to the students' daily routines have affect their performance. To the best of our knowledge, we believe that the presented dataset and results are the first to include both data from \textit{non-Covid-19} and \textit{Covid-19} periods. Furthermore, the model presented in this paper can predict the students' performance, even periods as different from one another as these.

The contributions of the paper can be summarised as follows: 
\begin{itemize}
    \item We present a new open-source dataset collected from a total of 61 students, from different subjects and in two different year.
    \item To the best of our knowledge the present dataset, is the first to join data from \textit{non-Covid-19} and \textit{Covid-19} periods of time.
    \item We present a model that is able to predict the students' performance with 80\% accuracy, using the data from one week.
    \item We present a pipeline that implements a final median voting mechanism that further improves the model's accuracy over time. 

\end{itemize}

The rest of the paper is organized as follows. Section \ref{sec:related_work} presents an overview of related works. Section \ref{sec:dataset} provides details on \gls{ISABELA} data acquisition and the used dataset. In section \ref{sec:model_results}, we present the results  using several models, for the 2018 and 2021 datasets, as well as for the joined dataset. The conclusion and guidelines for future work are presented in section \ref{sec:conclusion}.

\section{Related work}
\label{sec:related_work}

As stated before, in this paper we use the data collected in the trials performed with the \gls{ISABELA} application. In previous work we have explored how the data gathering process and the implementation of the system was performed. Additionally, in past publications, we covered an analysis of some of the collected data from two real world trials, namely, the analysis of the data collected Online Social Networks (i.e., Facebook) from students in Portugal, and the data collected from the use of the application by students in Ecuador \cite{fernandes2020isabela}. Furthermore, in the same work we also presented three different machine learning models that leveraged the collected data, namely a model to infer sleep periods, a model to classify sleep quality and a model to classify sociability. Additionally, a longitudinal analysis of some of the collected metrics  was also performed, namely, an analysis of the levels of activity, time spent studying, and the time spent in the university. In the current paper, we extend our previous work by proposing a new model to automatically infer students' performances, more specifically, a model capable of classifying the students belonging to the same school class into two groups, the students that perform \textit{above the median} and the students that perform \textit{below the median}.

Other studies have also proposed the use of smartphones as a tool to gather a variety of data on users and their environment. One example of such a study is the “StudentLife” study \cite{wang2014studentlife}, which used a smartphone application to collect data and infer students’ academic performance. The objective of this study was to correlate students' psychological and physical well-being, with their academic performance. The application collected students' activity levels, sleep quality, and conversations, as well as states such as stress and mood, through a mobile version of \gls{EMA}\cite{shiffman2008ecological}. A study was conducted in Dartmouth University, during a 10-week period with 48 students. Results from this study included correlations between the progression of the term and the students’ academic performance, self-assessed mental state, and behavioural trends. Additionally, their findings pointed to a significant inverse correlation between sleep duration and pre- and post-depression, which is line with existing literature and studies in the field. Additional correlations between the frequency of conversations and overall social activity with depressive states are also pointed out in this study.

The same team behind "StudentLife" built on their previous work , by proposing a regression model to predict students' \gls{GPA} \cite{wang2015smartgpa}. The proposed model used data retrieved from smartphones, such as activity data, conversational data, class attendance, and social interaction. In this study, the authors chose a Lasso (Least Absolute Shrinkage and Selection Operator) \cite{tibshirani1996regression} regularized linear regression model. The results obtained from their model have ±0.179 mean absolute error from the ground truth. The study shows that there are clear relations between the students’ behavioural patterns and respective grades. However, the focus on \gls{GPA} can be limiting since this metric can only be applied to ungraded students. Furthermore, this metric is not normative across different countries. Additionally, external factors can influence the results of students (e.g., the Covid-19 pandemic), and even between editions of the same school subject there can be a variety of different conditions that impact the overall performance of students (e.g., changes on the lecturer or evaluation methods). As such, we believe that a classification scheme that can cope with varying conditions while determining students' performance is needed.

The work by  Harari et al., \cite{harari2017patterns} used the dataset obtained in \cite{wang2014studentlife} to detect patterns in the behaviour of students and how they related to health problems (e.g., mental health, physical health). In this study, the results were presented for weekly time windows, and the dataset was treated as a population where the mean value represents all the individuals. The authors focused on two types of behaviour patterns, namely the physical activity behaviour patterns and the sociability behaviour patterns. The study found interesting correlations (negative and positive) between the term rhythm (e.g., deadlines, classes, mid-term etc.) and students’ mood, sociability, physical activity, and sleep periods. Stability estimates for weekly activity and sociability (i.e., correlation of activity/sociability levels between adjacent weeks) were also highly correlated, namely with a mean correlation of 0.66 for activity and a mean correlation of 0.72 for sociability. The authors also found a significant correlation between students' ethnicity and academic class, with their variation activity and sociability trajectories. This work presents several interesting correlations between students' behaviour and their daily routines. Furthermore, these behaviour patterns can be tied to the final  students' performance. However, in this work the authors do not propose the use of this data, nor founded correlations in order to help predict the academic performance of students.

Other works also tried to predict the students' performance by using machine learning models. This is the case of \cite{osmanbegovic2012data}. However, in this work, instead of smartphone data, the authors propose a model that uses demographic information, as well as past academic information, to predict the students' future performance. The authors present a dataset of 257 students and explore the use of three different models, namely a Naive Bayes algorithm, a Multilayer Perceptron, and the J48 algorithm, to classify student performance into one of two classes (students that fail, and students that pass). The results show that the tested models have an average recall of 85\% when detecting the students that pass. However, the models present very poor result when classifying the students that fail. The use of demographic information and past academic information can be interesting and valuable in this kind of model. However, this can also raise several privacy concerns, especially for a smaller dataset, since this type of information can be used to infer the students' identities.

The Covid-19 pandemic has also affected people's daily lives in several ways in the recent years, as several studies have pointed out \cite{afonso2020impact},\cite{buzzi2020psycho}. This is also true for students, which had their routines changed, were confined to their homes and had to change to a remote class system. In \cite{nepal2022covid}, a study was conducted to assess behavioural changes of 180 undergraduate college students, during the pandemic, using mobile phone sensing and behavioural inference. This study was divided in two acquisition phases. The first phase occurred one year prior to the pandemic and was used as baseline. The second phase occurred during the first year of the pandemic. The authors used \gls{PCA} to reduce the dimensionality of data and be able to cluster the students into different groups. Specifically, it was possible to identify two distinct groups. It was also observed that one of the groups was mostly composed of students with higher self-reported Covid-19 concern and higher levels of stress and anxiety. This shows that in fact different groups of students experience the pandemic differently. Additionally,  the authors found that that there was a positive correlation between levels of depression, anxiety and stress with self-reported Covid-19 concern. Additionally, a deep learning model was implemented to classify student Covid-19 concerns, with an \gls{AUC} score of 0.70 and a F1 score of 0.71. Mental health and stress can be directly correlated with student performance. As such, evaluating these metrics can be important. Furthermore, this study reveals interesting insights into students behaviour changes due to the Covid-19 pandemic. However, this study does not directly address the prediction of students' performance. 

\begin{figure*}[t!]
\centering
\begin{subfigure}{.30\textwidth}
\centering
\includegraphics[width=2in]{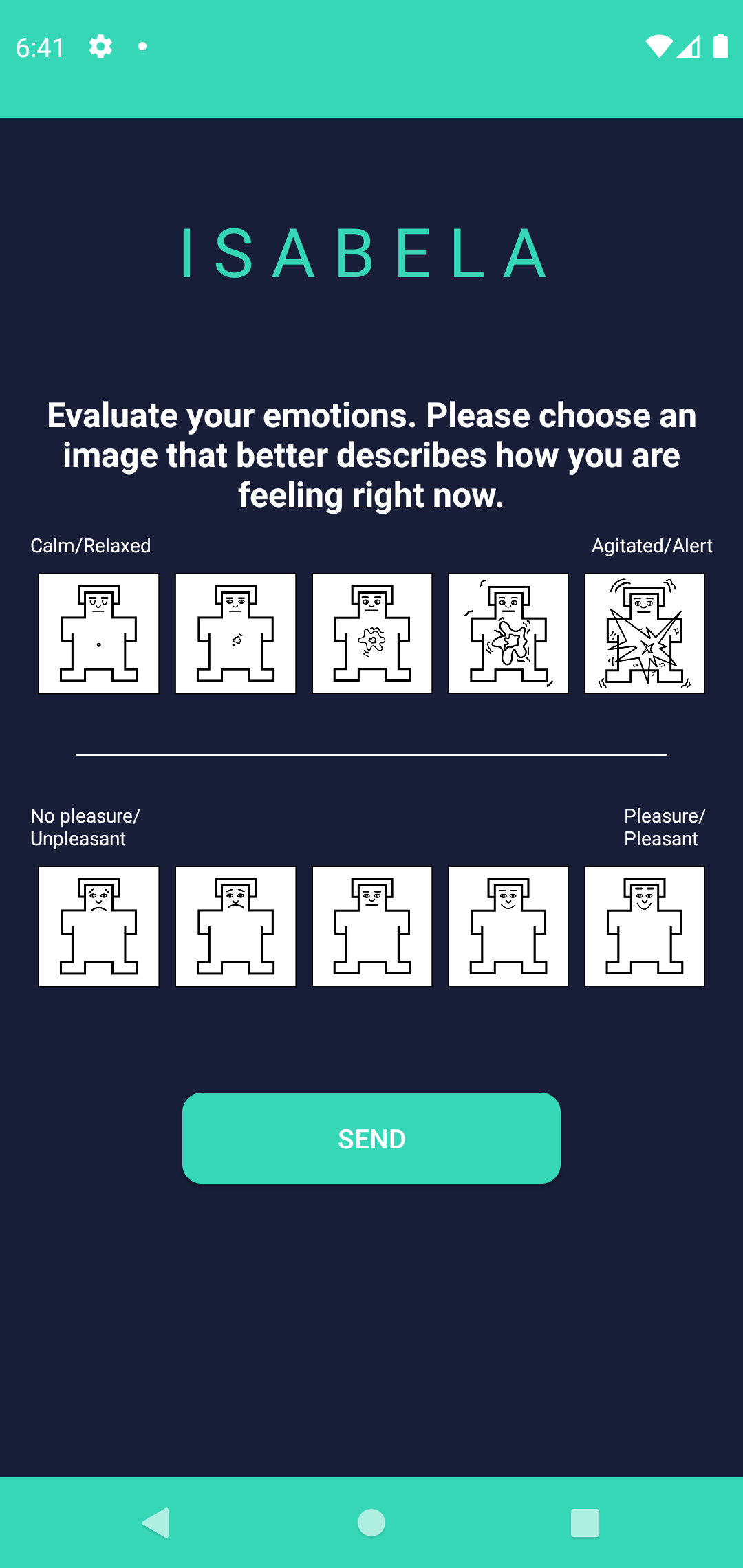}
\caption{}
\label{fig:emotional_form}
\end{subfigure}
\begin{subfigure}{.30\textwidth}
\centering
\includegraphics[width=2in]{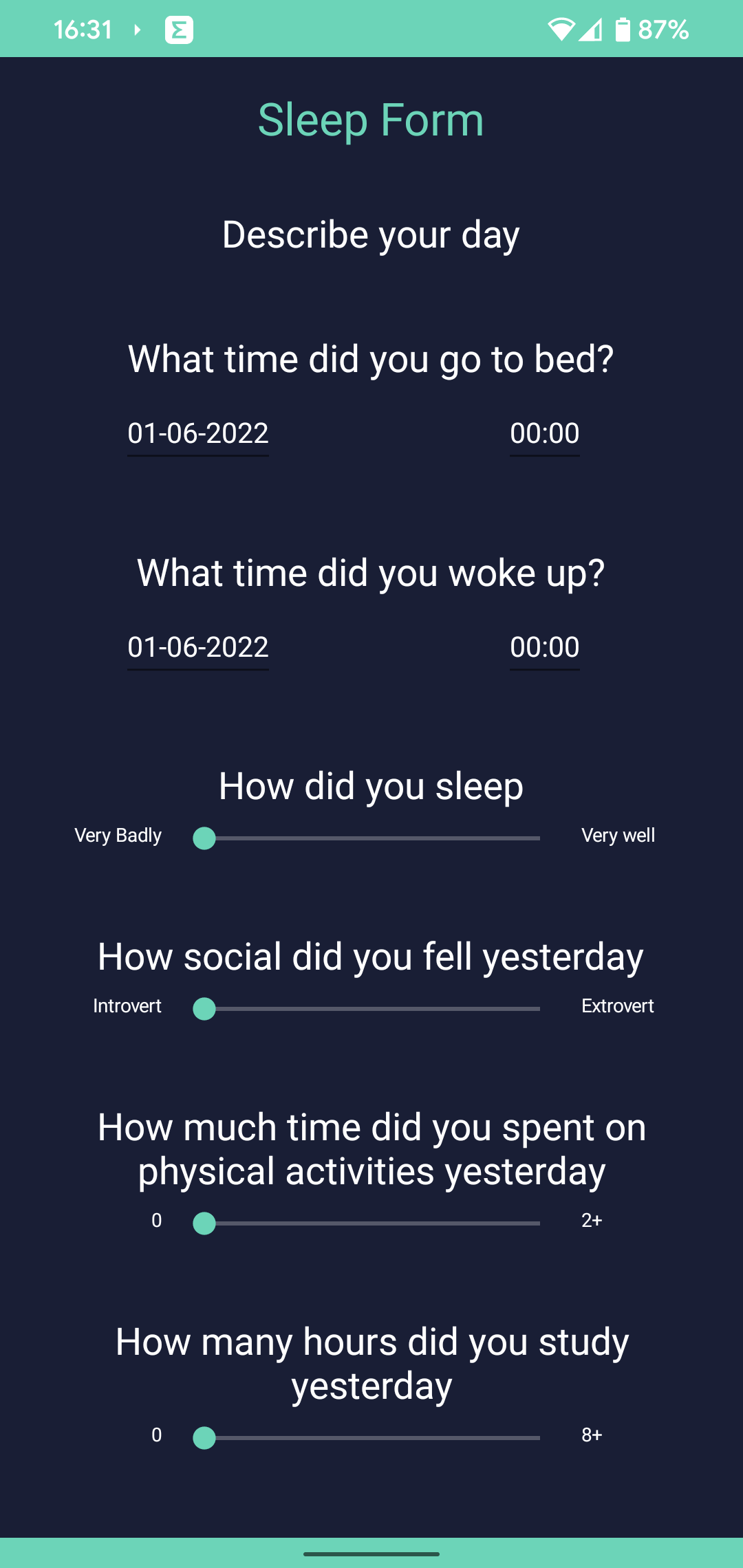}
\caption{}
\label{fig:sleep_form}
\end{subfigure}
\caption{\gls{ISABELA}'s Questionnaires: Emotional questionnaire using the \gls{SAM} scale (a); Application daily questionnaire (b);}
\label{fig:forms}
\end{figure*}

The early detection of students with poor performance can also be used to reduce the dropout rates. In \cite{sandoval2020early}, a study was conducted to create a model to predict which students were at risk of dropping out.  Additionally, this study is also relevant since it was carried out in the same academic institution, in which the \gls{ISABELA} datasets were acquired, namely the Escuela Politecnica Nacional, in Quito, Ecuador. The model uses students' demographic and academic information to classify students into one of two classes (students who dropout and students who do not). Furthermore, the authors tested two models, namely a Neural Network and a Logistic activation function. Although, the authors present a 76\% accuracy for the Neural Network model, the model presented had very high false positive rate. This is, due to the fact that the dataset is highly imbalanced (72.8\% Dropout, 27.2\% No Dropout) and the authors did not address that. However, the authors defend the results by saying that for their specific case study, it is better to raise alarms for interventions in students who do not need them (false positives) than to fail to identify students who do need them (false negatives). Additionally, it was possible to identify some patterns in students that are more probable to drop out, namely a low application grade, a low vulnerability index, enrolment in the month of March (the second of two available enrolment epochs), and being enrolled in bachelor' degrees. Although, this work does not address performance prediction, we believe that some of its findings can give some insights into the constraints and patterns of students from that university.

Although, there has been some work in the field of students' performance prediction, none of the concerned studies led to results that can be widely used independently of academic level or region/country dependent metrics (i.e., \gls{GPA}). Furthermore, as far as we know there is not any work that is able to predict the performance of students before and during the Covid-19 pandemic. As such, we believe that there is still a need for the creation of models that are able to automatically detect student performance in these scenarios.

\section{ISABELA Dataset}
\label{sec:dataset}

In our previous work in \cite{fernandes2020isabela} we build a system capable of collecting both passive and self-reported data from students. That system was then used to monitor students during two real world trials. In this section will cover the data acquisition procedure and the features of the datasets used in the models presented in this paper, as well as provide some additional information on the datasets, such as time of acquisition, demographics, and class labelling.  

\subsection{Data acquisition and features}
\label{sec:data_aquisition}

One of the main components of students' lives that the \gls{ISABELA} system aims to monitor, is their activity levels. The system  gathers this data by resorting to Google's Physical Activity Recognition \gls{API} \cite{activityRecognition}. This \gls{API} is able to classify user's activity into one of eight possible activities, namely: running, walking, on bicycle, in vehicle, on foot, tilting,  still, and unknown. 
The \gls{ISABELA} system receives an update every time the classification of the student's activity changes. Until a new classification is received the system considers the last valid classification as the ongoing activity. Additionally, since we are not interested in any particular type of exercise, but rather in activity/exercise levels, all of the categories that represent exercise were grouped into one category (\textit{"exercise"}). Thus, running, walking, on foot and on bicycle were grouped under the \textit{"exercise"} classification. Moreover, although \textit{tilting} does not represent any specific user activity, and is just an indicator of  smartphone tilting, we believe that this classification can also be used to infer the smartphone's usage.

The activity levels of students are one of the metrics that might have changed more due to the pandemic context. Due to sudden change in habits caused by confinement, most people saw their daily routines affected in terms of physical activity. As such, this feature can be important to study and characterise those periods of time. 

Another aspect that could be important to evaluate students' performance is their location. Using the students' location information, it is possible to infer if a student attended classes, spent time at home, or even if it frequently went out at night. Due to privacy concerns, our system did not store any GPS information. Instead, the system infers the discrete location of the user, namely by categorizing their location into one of three possible categories: \textit{home}, \textit{university}, or \textit{other}. The user configures the application on the first time it opens, by using the application capabilities to scan the available networks and select the SSID of his/her home network. The name of their home network is locally stored in a database on the smartphone. Due to privacy concerns this information never leaves the smartphone. The \gls{ISABELA} application scans available networks periodically and compares the results of the scans with the stored  SSID to infer if a user is at \textit{"home"} or not. All of the universities in Ecuador make the \textit{"Eduroam"} network available throughout the campuses. As such the application, searches for this particular network to determine if a user is at the \textit{"university"}. Lastly, if none of the two networks exist in the WiFi scan's result, the application categorizes the student's location as \textit{"other"}. 

Due to the lockdowns caused by the Covid-19 pandemic, location is also an important variable during this period of time. Additionally, students in Ecuador attended classes remotely during lockdown periods, thus eliminating the time spent at their university. Thus, to make both datasets compatible, the \textit{"university"} and \textit{"other"} discrete location categories were grouped together as \textit{"other"} for both datasets.

Although the smartphone's context information and the sensors data can be used to infer several daily life aspects, there are some aspects that are harder to detect trough passively collected data, especially those pertaining to  psychological states, since passive data has yet to be validated as an indicator for these kinds of metrics. As such, there is a need to use more conventional methods, such as questionnaires. In the \gls{ISABELA} system, we used questionnaires to complement the aforementioned passively collected data. 

One of the aspects, that can be relevant when evaluating student's performance is their emotional state. A questionnaire-based approach, based on the \gls{SAM} scale \cite{Bynion2020}, was implemented in order to obtain this type of data. This questionnaire prompted to the user on a daily base in order to obtain values of the self-perceived emotional states for each day. The \gls{SAM} scale is an image based self-assessment scale to evaluate the users' two-dimensional map of emotions. The questionnaire can be seen in Figure \ref{fig:emotional_form}. From top to bottom, the first scale corresponds to arousal while the second scale corresponds to valence. The user selects, one image from line, corresponding to an integer value ranging from 0 to 4, which indicates their valence and arousal values. These values can then be converted to a range of -1 to 1, with a step of 0.5, and be subsequently mapped to the 2-dimensional emotional map \cite{yazdani2013multimedia}. The questionnaire was prompted to the users at a randomly selected hour between 14-20h. Additionally, the questionnaire was not released during the morning since during this period the user's perception can still be affected by the events of the previous day.

Another type of data  that the \gls{ISABELA} system is able to collect, and that is highly relevant to the students' academic performance \cite{taras2005sleep}, is sleep data. This is done by prompting the user with a questionnaire every morning, which can be seen in Figure \ref{fig:sleep_form}. In this questionnaire, users can indicate the time at which they went to bed, the time at which they woke up, and how did they classify sleep quality. In addition, the users can also indicate their sociability level for the previous day, how many hours of exercise they had, and how many hours did they study.  

The \gls{ISABELA} system is able to collect other metrics that are not addressed in this paper. However, we believe that the previously presented data is the most relevant concerning students' performance. Additionally, some functionalities were added for the 2021 trial, while others were removed and, thus, not all collected metrics were presented in both datasets. A list of the selected features for this work are shown in the Table \ref{tab:features}. As it is possible to see in the table, the selected features are a mix of passively collected data from the smartphone's sensors/context and data that was self-reported by the users. We believe that the use of these two types of data makes the model more robust.

\subsection{Dataset information}
\label{sec:dataset_info}

The datasets presented in this paper were obtained in two distinct periods of time, and in two real world trials with the \gls{ISABELA} system. Furthermore, those trials were performed in different years and with different school subjects. As mentioned before the trials, were performed in Ecuador's Escuela Politecnica Nacional, in Quito. The first trial involved 28 students, from the 12$^{th}$ of May to the 12$^{th}$ of June, 2018. The second trial involved 33 students, from the 1$^{st}$ of March to the 31$^{st}$ of March, 2021.

All of the collected data was anonymized and, as such, the dataset does not include any personal information (e.g., age, gender, working status). For this reason, is not possible to determine the age, gender or other information for a specific user. However, from the school subject information, we known that both trials had students with different ages, genders, and working status. On the other hand, both sets of subjects were enrolled in the computer science degree, which normally includes a larger percentage of male students. All students had Ecuadorian nationality. Nonetheless, we believe that the dataset is representative of a larger population.

\begin{table}[t!]
\centering
\renewcommand{\arraystretch}{2}
\caption{Dataset's Features }
\label{tab:features}
\begin{tabular}{|c|c|}
\hline
                                              & \textbf{Feature}                      \\ \hline
\multirow{7}{*}{\rotatebox[origin=c]{45}{\textbf{Sensors and Context}}} & \% of time at \textit{other}          \\ \cline{2-2} 
                                              & \% of time at \textit{house}          \\ \cline{2-2} 
                                              & \% of time \textit{still}             \\ \cline{2-2} 
                                              & \% of time \textit{exercise}          \\ \cline{2-2} 
                                              & \% of time \textit{in vehicle}        \\ \cline{2-2} 
                                              & \% of time \textit{unknown}           \\ \cline{2-2} 
                                              & \% of time \textit{tilting}           \\ \hline
\multirow{7}{*}{\rotatebox[origin=c]{45}{\textbf{Surveys}}}             & self-reported arousal levels          \\ \cline{2-2} 
                                              & self-reported valence levels          \\ \cline{2-2} 
                                              & self-reported sociability levels      \\ \cline{2-2} 
                                              & self-reported sleep quality levels    \\ \cline{2-2} 
                                              & self-reported sleep hours             \\ \cline{2-2} 
                                              & self-reported physical activity hours \\ \cline{2-2} 
                                              & self-reported study hours             \\ \hline
\end{tabular}
\end{table}

The 2021 dataset was obtained during the lockdown caused by the Covid-19 pandemic. During the lockdowns students were confined to their homes, which could lead to lower exercise levels, did not have as much physical and/or social interactions, did not have the option to do leisure activities outside, and had to attend classes remotely. These changes had the potential to cause negative effects not only on the students' physical well-being, but also on their psychological and social wellbeing\cite{haleem2020effects}. Additionally, these changes had the potential to affect the performance of students. Because the trial periods can be so different in nature (as is the case with the 2018 and 2021 trials), it is important that any model for automatically predicting students' performance is able to deal with data samples from heterogeneous periods.

\begin{table}[t]
\renewcommand{\arraystretch}{2}
\caption{Final grades metrics in the 2018 and 2021 datasets.}
\label{tab:grade_metrics}
\begin{tabular}{|c|c|c|c|c|c|}
\hline
                       & \textbf{Min} & \textbf{Max} & \textbf{Median} & \textbf{Approved} & \textbf{Non-Approved} \\ \hline
\textit{\textbf{2018}} & 9.6          & 14.4         & 12.6            & 20                & 8                 \\ \hline
\textit{\textbf{2021}} & 12.14        & 17.72        & 14.90           & 33                & 0                 \\ \hline
\end{tabular}
\end{table}

\begin{table*}[t!]
\centering
\renewcommand{\arraystretch}{2}
\caption{Number of students and sample size by dataset and time windows.}
\label{tab:number_samples}
\begin{tabular}{|c|c|ccc|ccc|}
\hline
\multirow{2}{*}{} & \multirow{2}{*}{\textbf{Number of Students}} & \multicolumn{3}{c|}{\textbf{Monthly data aggregation}}                                                                                     & \multicolumn{3}{c|}{\textbf{Weekly data aggregation}}                                                                                      \\ \cline{3-8} 
                  &                                              & \multicolumn{1}{c|}{\textit{\textbf{Below Median}}} & \multicolumn{1}{c|}{\textit{\textbf{Above Median}}} & \textit{\textbf{Total}} & \multicolumn{1}{c|}{\textit{\textbf{Below Median}}} & \multicolumn{1}{c|}{\textit{\textbf{Above Median}}} & \textit{\textbf{Total}} \\ \hline
\textbf{2018}     & 28                                           & \multicolumn{1}{c|}{15}                             & \multicolumn{1}{c|}{13}                            & 28                      & \multicolumn{1}{c|}{65}                             & \multicolumn{1}{c|}{53}                            & 118                     \\ \hline
\textbf{2021}     & 33                                           & \multicolumn{1}{c|}{15}                             & \multicolumn{1}{c|}{18}                            & 33                      & \multicolumn{1}{c|}{38}                             & \multicolumn{1}{c|}{44}                            & 82                      \\ \hline
\textbf{Joined}   & 61                                           & \multicolumn{1}{c|}{30}                             & \multicolumn{1}{c|}{31}                            & 61                      & \multicolumn{1}{c|}{103}                            & \multicolumn{1}{c|}{97}                            & 200                     \\ \hline
\end{tabular}
\end{table*}

The students' final grades in the school subjects, for each trial, were also considered, to be used as classification label. Only the professor of the subject was aware of student's identities. Thus, we only obtained the final grade of the subject, for each corresponding anonymous id of the \gls{ISABELA} application, hence students' anonymity was maintained.  This metric was selected, instead of their average final classification, because not all students were enlisted at the same school subjects. The school subjects that one takes could influence their average final classification, and as such this metric would not allow for a direct comparison between students. In Table \ref{tab:grade_metrics}, we can see the minimum, maximum and median grade of each dataset, as well as the number of students that obtained a passing grade and the number of students that did not. In these school subjects, a student had to have a minimum grade of 12 points, to be approved. 

We can see from the table that the distribution of the students' grades is quite different for both trials. Furthermore, in the 2021 dataset not only all students were approved, but also the median grade is higher than the maximum grade obtained in the 2018 trial. Due to these aspects, the labelling of each dataset was made separately, even when the datasets were jointly considered. The students were divided into two groups to perform a binary classification, namely the ones that obtained a grade \textit{"above the median"} value (True class) and the ones that obtained a grade \textit{"below the median"} value (False class). Even among editions of the same school subject for different years, we can have different levels of difficulty due to the changes in the lecturing or evaluating process. Using the median value allows us to create a model that is able to predict the best and lowest performance students among a class, even considering disparities between school subjects or editions. 

We can see the final distribution of the dataset's samples in the two classes in Table \ref{tab:number_samples}. Additionally, in Table \ref{tab:number_samples}, we present the distribution of samples considering two data aggregation periods, namely monthly data aggregation and weekly data aggregation.  We will explore the implications of using different data aggregation time windows, in section \ref{sec:individual_datasets}.

\section{Models and Results}
\label{sec:model_results}

Determining which students will do worse or better within a class can be useful information, to reduce dropout rates and take measures to increase the overall performance of the class. In this section we explore how the data acquired from both trials can be used to create a model for this purpose. Additionally, in this section we present a final classification vote mechanism that improves the general performance of our model and can make the model increase its performance over time. 

All the tests presented below were performed with \gls{LOOCV}. This particular type of cross-validation uses a number of folds that is equal to the number of samples of the dataset, that is, all but one sample are used as training set and the remaining instance is used as a single-item test set for each run. This allows us to maximize the training set of each cross-validation fold, which also allows the creation of a more generic model.

All the code, for the results presented below, was produced using python and the scikit-learn library. The used python version was 3.7.6, and the scikit-learn version was 1.0.2. 

\subsection{Individual Datasets}
\label{sec:individual_datasets}

\begin{table*}[t]
\caption{Performance of distinct model by time interval, for the 2018 dataset.}
\label{tab:time_vs_performance_2019}
\centering
\renewcommand{\arraystretch}{2}
\begin{tabular}{|c|ccc|ccc|}
\hline
\multirow{2}{*}{}                                                                  & \multicolumn{3}{c|}{\textbf{Monthly data aggregation}}                                                                                          & \multicolumn{3}{c|}{\textbf{Weekly data aggregation}}                                                                                           \\ \cline{2-7} 
                                                                                   & \multicolumn{1}{c|}{\textbf{sensitivity}} & \multicolumn{1}{c|}{\textbf{specificity}} & \multicolumn{1}{c|}{\textbf{accuracy}} & \multicolumn{1}{c|}{\textbf{sensitivity}} & \multicolumn{1}{c|}{\textbf{specificity}} & \multicolumn{1}{c|}{\textbf{accuracy}} \\ \hline
\textbf{Decision Tree}                                                             & \multicolumn{1}{c|}{0.492}              & \multicolumn{1}{c|}{0.567}                & 0.532                                  & \multicolumn{1}{c|}{0.606}              & \multicolumn{1}{c|}{0.705}                & 0.660                                  \\ \hline
\textbf{Random Forest}                                                             & \multicolumn{1}{c|}{0.477}              & \multicolumn{1}{c|}{0.607}                & 0.546                                  & \multicolumn{1}{c|}{0.715}              & \multicolumn{1}{c|}{0.794}                & 0.758                                  \\ \hline
\textbf{SVM}                                                                       & \multicolumn{1}{c|}{0.385}              & \multicolumn{1}{c|}{0.533}                & 0.464                                  & \multicolumn{1}{c|}{0.736}              & \multicolumn{1}{c|}{0.846}                & 0.797                                  \\ \hline
\textbf{Naive Bayes}                                                               & \multicolumn{1}{c|}{0.692}              & \multicolumn{1}{c|}{0.533}                & 0.607                                  & \multicolumn{1}{c|}{0.811}              & \multicolumn{1}{c|}{0.646}                & 0.720                                  \\ \hline
\textbf{K-Near-Neighbours}                                                         & \multicolumn{1}{c|}{0.846}              & \multicolumn{1}{c|}{0.267}                & 0.536                                  & \multicolumn{1}{c|}{0.585}              & \multicolumn{1}{c|}{0.877}                & 0.746                                  \\ \hline
\textbf{\begin{tabular}[c]{@{}c@{}}AdaBoost-SAMME w/\\ Random Forest\end{tabular}} & \multicolumn{1}{c|}{0.569}              & \multicolumn{1}{c|}{0.6}                  & 0.586                                  & \multicolumn{1}{c|}{0.751}              & \multicolumn{1}{c|}{0.777}                & 0.765                                  \\ \hline
\textbf{XGBoost}                                                                   & \multicolumn{1}{c|}{0.769}              & \multicolumn{1}{c|}{0.533}                & 0.643                                  & \multicolumn{1}{c|}{0.755}              & \multicolumn{1}{c|}{0.815}                & 0.788                                  \\ \hline
\end{tabular}
\end{table*}

\begin{table*}[]
\caption{Performance of distinct model by time interval, for the 2021 dataset}
\label{tab:time_vs_performance_2021}
\centering
\renewcommand{\arraystretch}{2}

\begin{tabular}{|c|ccc|ccc|}
\hline
\multirow{2}{*}{}                                                                  & \multicolumn{3}{c|}{\textbf{Monthly data aggregation}}                                                                                          & \multicolumn{3}{c|}{\textbf{Weekly data aggregation}}                                                                                           \\ \cline{2-7} 
                                                                                   & \multicolumn{1}{c|}{\textbf{sensitivity}} & \multicolumn{1}{c|}{\textbf{specificity}} & \multicolumn{1}{c|}{\textbf{accuracy}} & \multicolumn{1}{c|}{\textbf{sensitivity}} & \multicolumn{1}{c|}{\textbf{specificity}} & \multicolumn{1}{c|}{\textbf{accuracy}} \\ \hline
\textbf{Decision Tree}                                                             & \multicolumn{1}{c|}{0.483}              & \multicolumn{1}{c|}{0.433}                & 0.461                                  & \multicolumn{1}{c|}{0.777}              & \multicolumn{1}{c|}{0.537}                & 0.666                                  \\ \hline
\textbf{Random Forest}                                                             & \multicolumn{1}{c|}{0.528}              & \multicolumn{1}{c|}{0.407}                & 0.473                                  & \multicolumn{1}{c|}{0.791}              & \multicolumn{1}{c|}{0.718}                & 0.757                                  \\ \hline
\textbf{SVM}                                                                       & \multicolumn{1}{c|}{0.611}              & \multicolumn{1}{c|}{0.600}                & 0.606                                  & \multicolumn{1}{c|}{0.864}              & \multicolumn{1}{c|}{0.711}                & 0.793                                  \\ \hline
\textbf{Naive Bayes}                                                               & \multicolumn{1}{c|}{0.722}              & \multicolumn{1}{c|}{0.333}                & 0.545                                  & \multicolumn{1}{c|}{0.841}              & \multicolumn{1}{c|}{0.605}                & 0.732                                  \\ \hline
\textbf{K-Near-Neighbours}                                                         & \multicolumn{1}{c|}{0.444}              & \multicolumn{1}{c|}{0.733}                & 0.576                                  & \multicolumn{1}{c|}{0.864}              & \multicolumn{1}{c|}{0.605}                & 0.744                                  \\ \hline
\textbf{\begin{tabular}[c]{@{}c@{}}AdaBoost-SAMME w/\\ Random Forest\end{tabular}} & \multicolumn{1}{c|}{0.678}              & \multicolumn{1}{c|}{0.533}                & 0.612                                  & \multicolumn{1}{c|}{0.814}              & \multicolumn{1}{c|}{0.745}                & 0.782                                  \\ \hline
\textbf{XGBoost}                                                                   & \multicolumn{1}{c|}{0.611}              & \multicolumn{1}{c|}{0.467}                & 0.545                                  & \multicolumn{1}{c|}{0.750}              & \multicolumn{1}{c|}{0.684}                & 0.720                                  \\ \hline
\end{tabular}
\end{table*}

As previously stated, the total available dataset is made up of two distinct trials, in two distinct years, namely 2018 and 2021. In both trials students participated for one month each. Furthermore, due to the Covid-19 pandemic the datasets are constituted by data from a period when students had in person classes (i.e., 2018 trial) and data from a period when classes were fully remote (i.e., 2021 trial). As such, the datasets from both trials are intrinsically very different. This is particularly true since we are using both the activity levels and the location of students as features for our models. For these reasons, we decided to start by evaluating the performance of each of the considered models when applied to each individual dataset.

Furthermore, we compare the results obtained by each model considering the data aggregated by student and by different intervals of time, namely the data aggregated by month and by week. Each feature is firstly aggregated by the mean value per day and per user, and then the mean value of each feature is calculated for each month and week respectively. The results for the various models, and for the different aggregation intervals are showed in Tables \ref{tab:time_vs_performance_2019} and \ref{tab:time_vs_performance_2021}, for the 2018 and 2021 datasets, respectively.

Seven different models were tested, namely: Decision Tree\cite{quinlan1996learning}; Random Forest\cite{breiman2001random}; \gls{SVM}\cite{mammone2009support}; Naive Bayes\cite{rish2001empirical}; K-Near-Neighbours\cite{cover1967nearest}; AdaBoost-SAMME\cite{hastie2009multi} with Random Forest (hereafter referred to as AdaBoost); and XGBoost\cite{chen2016xgboost}. All of the models shown in Table \ref{tab:time_vs_performance_2019} and \ref{tab:time_vs_performance_2021} were obtained by running the default implementation of scikit-learn, and all models were trained by using \gls{LOOCV}. Using  this method removes any variance that normally is created by the random distribution of the samples by folds. However, each model has a random start state, which can lead to different results. For that reason, every model was run 10 times and the mean values for sensitivity, specificity and accuracy were calculated. Furthermore, a grid-search was performed for each model and on each dataset, to find the best parameters for each case. The grid search space as well as the best configuration of each model can be seen in the appendix. 

As we can see in the tables, in both datasets, the models that use data aggregated by week surpass the models that use the data aggregated by month. This is partly due to the smaller number of samples available to the models, when data is aggregated by month, as we show in Table \ref{tab:number_samples}. However, as we will see in the next section, even when the sample size increases, the weekly aggregation models still surpass the ones that use monthly aggregation. We believe that the reason behind this is the levelling out of changes in the data when aggregating the data over long periods. For example, if a student skips school one day every other week, by month that would represent 6\% of the time, while by week it would represent more than the double (15\%) of the time. We believe that a temporal window of a week to aggregate the data, is the best approach since it can better grasp sudden variations in the student's behaviour. As such, for the remaining of the tests, presented in this section, we will focus on this aggregation window.

\begin{table}[h!]
\caption{Cost-Sensitive Performance of SVM and AdaBoost for both datasets, in the week time window.}
\label{tab:cost_sensitive}
\centering
\renewcommand{\arraystretch}{2}
\begin{tabular}{|cc|c|c|c|}
\hline
\multicolumn{2}{|c|}{}                                                   & \textit{\textbf{sensitivity}} & \textit{\textbf{Specificity}} & \textit{\textbf{Accuracy}} \\ \hline
\multicolumn{1}{|c|}{\multirow{2}{*}{\textbf{2018}}} & \textbf{AdaBoost} & 0.792                         & 0.800                         & 0.797                      \\ \cline{2-5} 
\multicolumn{1}{|c|}{}                               & \textbf{SVM}      & 0.774                         & 0.800                         & 0.788                      \\ \hline
\multicolumn{1}{|c|}{\multirow{2}{*}{\textbf{2021}}} & \textbf{AdaBoost} &  0.795                        & 0.763                         &   0.780                       \\ \cline{2-5} 
\multicolumn{1}{|c|}{}                               & \textbf{SVM}      & 0.841                         & 0.711                           & 0.780                      \\ \hline
\end{tabular}
\end{table}

As mentioned before, both datasets are only slightly imbalanced and, as such we can use accuracy as metric to evaluate the models' performance. We can see that the models which present the best performance in both tables are the \gls{SVM} and the AdaBoost models. The AdaBoost model offers a performance improvement over the Random Forest model, with only a slight improvement in the 2018 dataset, but a bigger improvement in the 2021 dataset. Furthermore, in both cases, the \gls{SVM} and AdaBoost models obtained an accuracy higher than 76\%. 

\begin{table}[]
\caption{Cross testing with datasets.}
\label{tab:cross_testing}
\centering
\renewcommand{\arraystretch}{2}
\begin{tabular}{|cc|c|c|c|}
\hline
\multicolumn{2}{|c|}{}                                                     & \textit{\textbf{sensitivity}} & \textit{\textbf{Specificity}} & \textit{\textbf{Accuracy}} \\ \hline
\multicolumn{1}{|c|}{\multirow{3}{*}{\textbf{Case A}}} & \textbf{AdaBoost} & 0.684                         & 0.750                         & 0.720                      \\ \cline{2-5} 
\multicolumn{1}{|c|}{}                                 & \textbf{SVM}      & 0.711                         & 0.750                         & 0.732                      \\ \cline{2-5} 
\multicolumn{1}{|c|}{}                                 & \textbf{C-S-AB}      & 0.711                         & 0.682                         & 0.695                      \\ \cline{2-5} 
\multicolumn{1}{|c|}{}                                 & \textbf{C-S-SVM}  & 0.763                         & 0.841                         & 0.805                      \\ \hline
\multicolumn{1}{|c|}{\multirow{3}{*}{\textbf{Case B}}} & \textbf{AdaBoost} & 0.642                         & 0.692                         & 0.669                      \\ \cline{2-5} 
\multicolumn{1}{|c|}{}                                 & \textbf{SVM}      & 0.736                         & 0.692                         & 0.712                      \\ \cline{2-5} 
\multicolumn{1}{|c|}{}                                 & \textbf{C-S-AB}      & 0.679                         & 0.738                         & 0.712                      \\ \cline{2-5} 
\multicolumn{1}{|c|}{}                                 & \textbf{C-S-SVM}  & 0.755                         & 0.569                         & 0.653                      \\ \hline
\end{tabular}
\end{table}

However, we can also see from the tables that the best two models have distinct behaviours, the AdaBoost model has a more balanced distribution of sensitivity and specificity in both datasets, while, the \gls{SVM} models tends to favour the class with majority of samples in both datasets. For that reason, we also used Cost-Sensitive learning to evaluate if the class imbalance was affecting the model's final performance. For this particular test, we only considered the weekly aggregation interval, since it shows better performance than the monthly aggregation approach. Grid search was used to find the best cost weight for each class in both datasets. Both \gls{SVM} and AdaBoost were tested with cost-sensitive learning and with cost weight of 1.125 for the minority class of each dataset, namely \textit{Above Median} in the 2018 dataset and \textit{Below Median} in the 2021 dataset. As we can see in the Table \ref{tab:cost_sensitive},  for the 2018 dataset the SVM model lost some of its accuracy, while the AdaBoost model increased its. The AdaBoost model increased its sensitivity by 4.1\% and its specificity by 2.3\%, leading to an increase of 3.2\% in accuracy. On the other hand, The \gls{SVM} model increased its sensitivity by 3.8\%, but had a decrease in its specificity by 4.6\%, which led to a less accurate model. However, this can still be a valid strategy, if the main purpose of the desired model is the correct detection of students that are \textit{above the median} value. In the case of the 2021 dataset, both models lost some of their accuracy. However, only the AdaBoost model had a trade-off between the sensitivity and specificity metrics, while the \gls{SVM} model lost accuracy in the majority class but did not gain any performance in the minority class. This shows that, in the particular case of AdaBoost, cost-sensitive learning can be used to improve the model performance. Furthermore, applying cost-sensitive learning shows better results in the case of the 2018 dataset. This could be due to  the difference in the ratio of the classes being bigger when compared to the 2021 dataset. 

Lastly, we also wanted to find out the answer to the question "can a model trained with one dataset be used to predict the performance of the other dataset?". This question is particularly relevant due to the difference in students' daily routines caused by the Covid-19 paradigm. For this particular test, once again we only considered the weekly aggregation approach and the models with the best performance, namely AdaBoost, SVM, Cost-Sensitive-AdaBoost (C-S-AB) and Cost-Sensitive-SVM(C-S-SVM). We considered two particular cases, namely: Case \textit{A}, use the 2018 dataset (non-covid dataset) as training set and the2021 dataset (covid dataset) as testing set; and Case \textit{B}, use the 2021 dataset as the training set and the 2018 dataset as testing set. 

The results for both cases can be seen in Table \ref{tab:cross_testing}. We can see that case \textit{A} provides models with better performance. In particular, C-S-SVM led to an accuracy of over 80\%, which is higher than the one obtained when using \gls{LOOCV} to train and test with only the 2018 dataset. This could be due to the fact that the 2018 dataset has more samples and, as such, is able to generate a more generic model. Another reason that possibly contributes to this difference in accuracy, is the changes in lifestyle of students, caused by the Covid-19 pandemic. Several aspects of the students' daily routines were affected during this period, namely the confinement to their household, the decrease in exercise levels, changes in sleep patterns, and more. Because of that some variance between the trial subjects may have been lost and, as a result, the model trained with data from this period turned out to be less generic and representative of other periods. This may lead to its lower performance when evaluating samples prior to the 2021 dataset. However, in order to support these conclusions, tests with bigger datasets from both periods (outside and during Covid times) are needed.

\subsection{Joined Dataset}

\begin{table*}[t!]
\caption{Performance of distinct models by time interval, for the joined dataset.}
\label{tab:results_joined}
\centering
\renewcommand{\arraystretch}{2}
\begin{tabular}{|c|ccc|ccc|}
\hline
\multirow{2}{*}{}                                                                  & \multicolumn{3}{c|}{\textbf{Monthly data aggregation}}                                                                       & \multicolumn{3}{c|}{\textbf{Weekly data aggregation}}                                                                        \\ \cline{2-7} 
                                                                                   & \multicolumn{1}{c|}{\textbf{sensitivity}} & \multicolumn{1}{c|}{\textbf{specificity}} & \textbf{accuracy} & \multicolumn{1}{c|}{\textbf{sensitivity}} & \multicolumn{1}{c|}{\textbf{specificity}} & \textbf{accuracy} \\ \hline
\textbf{Decision Tree}                                                             & \multicolumn{1}{c|}{0.526}                & \multicolumn{1}{c|}{0.490}                & 0.508             & \multicolumn{1}{c|}{0.664}                & \multicolumn{1}{c|}{0.692}                & 0.678             \\ \hline
\textbf{Random Forest}                                                             & \multicolumn{1}{c|}{0.568}                & \multicolumn{1}{c|}{0.54}                 & 0.554             & \multicolumn{1}{c|}{0.732}                & \multicolumn{1}{c|}{0.717}                & 0.724             \\ \hline
\textbf{SVM}                                                                       & \multicolumn{1}{c|}{0.710}                & \multicolumn{1}{c|}{0.467}                & 0.590             & \multicolumn{1}{c|}{0.825}                & \multicolumn{1}{c|}{0.748}                & 0.785             \\ \hline
\textbf{Naive Bayes}                                                               & \multicolumn{1}{c|}{0.710}                & \multicolumn{1}{c|}{0.567}                & 0.639             & \multicolumn{1}{c|}{0.763}                & \multicolumn{1}{c|}{0.621}                & 0.690             \\ \hline
\textbf{K-Near-Neighbours}                                                         & \multicolumn{1}{c|}{0.677}                & \multicolumn{1}{c|}{0.300}                & 0.492             & \multicolumn{1}{c|}{0.773}                & \multicolumn{1}{c|}{0.612}                & 0.690             \\ \hline
\textbf{\begin{tabular}[c]{@{}c@{}}AdaBoost-SAMME w/\\ Random Forest\end{tabular}} & \multicolumn{1}{c|}{0.619}                & \multicolumn{1}{c|}{0.587}                & 0.603             & \multicolumn{1}{c|}{0.826}                & \multicolumn{1}{c|}{0.760}                & 0.792             \\ \hline
\textbf{XGBoost}                                                                   & \multicolumn{1}{c|}{0.645}                & \multicolumn{1}{c|}{0.467}                & 0.557             & \multicolumn{1}{c|}{0.794}                & \multicolumn{1}{c|}{0.777}                & 0.785             \\ \hline
\end{tabular}
\end{table*}

\begin{figure}[!t]
\centering
\includegraphics[width=3.3in]{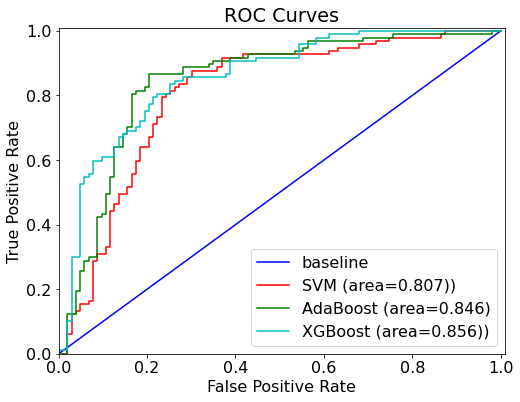}
\caption{ROC curve graph for the best performing models, namely SVM, AdaBoost and XGBoost.}
\label{fig:roc_curve}
\end{figure}

\begin{figure}[!t]
\centering
\includegraphics[width=3.3in]{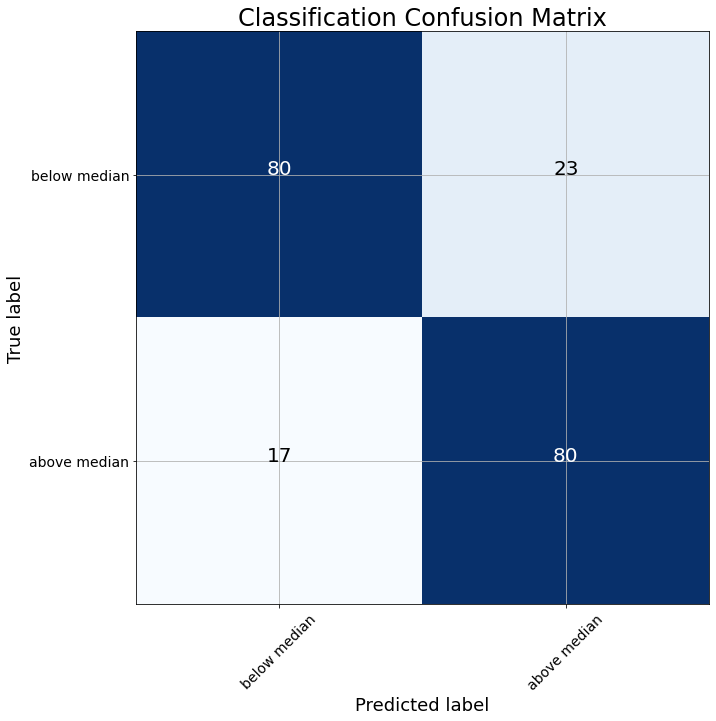}
\caption{Confusion Matrix AdaBoost-SAMME with Random Forest model.}
\label{fig:confuse_matrix}
\end{figure}

In the previous section we saw, that training with one dataset and testing with the other one can lead to mixed results. The other aspect that we were interested in evaluating was if both datasets, although having been collected in two different periods, could be used to create a more generic model. In fact, by joining both datasets, we increase the number of samples and this may lead to better evaluating the proposed models. After joining the datasets, the final dataset became almost perfectly balanced. As such, in this section we do not test cost-sensitive learning. As stated before, the aim of this work is to create a model able to predict students' performance, considering they are under the same conditions. However, even considering the same school subject, different editions (i.e., same school subject in different years) might be affected by changes in curriculum, teaching method, or even evaluation procedure. As such, class labelling was performed prior to joining the datasets, by computing each dataset grade median value separately and using them to label each dataset separately. As such, the class distribution is the sum of the students from each dataset in each class, presented in Table \ref{tab:number_samples}. 

The results for the various models being tested, for the joined dataset, considering both time aggregation approaches can be seen in Table \ref{tab:results_joined}. As in the case of the previous tests, a grid search was performed in order to obtain the optimal parameters for each model. In Table \ref{tab:results_joined} we can see that, similarly to what happened with the individual datasets, the models trained with the weekly aggregation approach outperform the ones trained with the monthly aggregation one. Once again, this might be because of the smaller sample size of the dataset with the monthly aggregation approach. However, as we stated previously, this seems to be only part of the reason, since even after doubling the sample size (by joining both datasets), most models see almost no performance increase, and some actually lose performance. This support our thesis that a month might be too big a time window, and some of the changes in students' behaviour might be ignored due to that. However, a bigger dataset is needed to validate this hypothesis. 

Considering the results obtained with the weekly aggregation, we can see that similarly, to what happened when working with the individual datasets, the models with the best performance are \gls{SVM}, AdaBoost and XGBoost. The AdaBoost model is the one that offers a higher accuracy, reaching almost 80\% of accuracy. We can also see the confusion matrix for the AdaBoost model in Figure \ref{fig:confuse_matrix}, where it is apparent that this model is able to predict correctly the class of 160 samples from the dataset. Additionally, the \gls{SVM} model and the XGBoost model have the same accuracy of 78.5\%. However, XGBoost offers a higher specificity, that is it is able to detect the students that are \textit{below the median} more accurately.

We can see the \gls{ROC} curves of each model in Figure \ref{fig:roc_curve}. Even though the AdaBoost model offers a higher accuracy, the XGBoost offers a \gls{AUC} score of 0.856 score, against 0.846 of the AdaBoost model and 0.807 of the \gls{SVM} model. This means that the XGBoost model is the model that is less prone to false positives. As such, the decision between one of these two models could depend on the particular use-case. That is, if we want to accurately determine both classes, XGBoost is the most suited model. However, if, for a particular case study, we wanted only to determine the students that are the best performers within a subject, AdaBoost would be the best model for that use case.    

The performed tests show that the AdaBoost model, the XGBoost model or even the \gls{SVM} model can be used to accurately predict which students will perform above the median and which students will perform under that threshold. Furthermore, the tests indicate that a model trained with the data from both datasets is able to obtain roughly the same performance as a model trained with each individual dataset. This also proves that is possible to create a model that is able to predict the performance of students with data collected in different periods, in this case before and during the Covid-19 pandemic.

\subsection{Median Classification Vote}

\begin{figure*}[!t]
\centering
\includegraphics[width=7in]{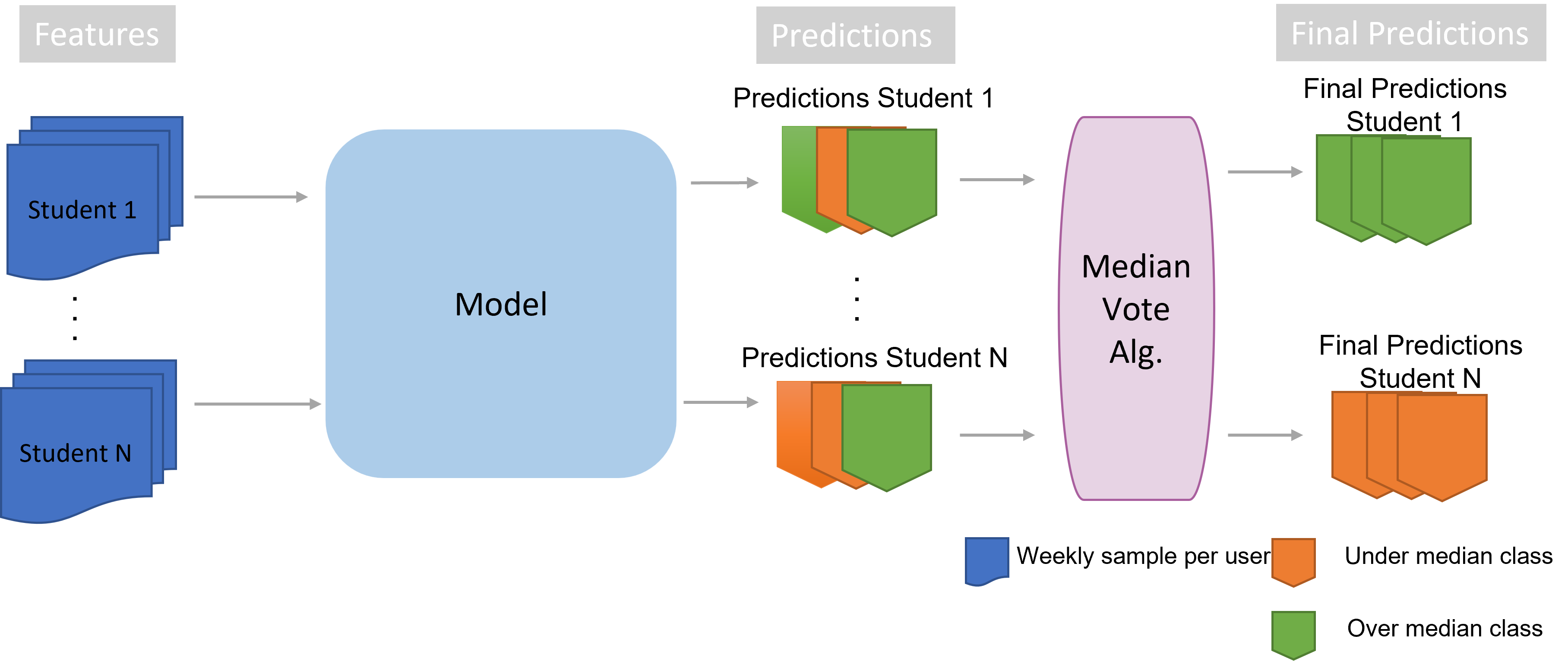}
\caption{Diagram of classification using proposed the median voting algorithm to obtain the final classification.}
\label{fig:pipeline_median}
\end{figure*}

\begin{algorithm}[t!]
\caption{Median vote of classifications.}\label{alg:median_vote}
\begin{algorithmic}
\Require 
\State $P \gets$ Array of Predictions 
\State $S \gets$ Array of Student identifiers for each prediction
\\
\Output{$FinalPredictions$}
\\
\Ensure $Size(P) == Size(S)$
\\
\State $P(s) \gets $  subset of Predictions for student s
\\

\State $FinalPredictions \gets Copy(P)$ 
\\
\State $i \gets 0$
\\

\While{$i < Size(P) $}
    \\
    \State $s \gets S[i]$
    \\
    \State $p \gets  Median(P(s))$
    \\
    \If{$p \geq 0.5$}
    \\
        \State $FinalPredictions[i] \gets 1$
    \Else{}
    \\
        \State $FinalPredictions[i] \gets 0$
    \EndIf
    \\
    \State $ i \gets i+1$
    \\
\EndWhile

 \end{algorithmic}
\end{algorithm}

\begin{table*}[h!]
\centering
\renewcommand{\arraystretch}{2}
\caption{Performance of Models after using the Median Voting Algorithm at decision level.}
\label{tab:performance_pipeline}
\begin{tabular}{|c|ccc|ccc|}
\hline
\multirow{2}{*}{}                                                                  & \multicolumn{3}{c|}{\textbf{Weekly data aggregation}}                                                                        & \multicolumn{3}{c|}{\textbf{By Student}}                                                                  \\ \cline{2-7} 
                                                                                   & \multicolumn{1}{c|}{\textbf{sensitivity}} & \multicolumn{1}{c|}{\textbf{specificity}} & \textbf{accuracy} & \multicolumn{1}{c|}{\textbf{sensitivity}} & \multicolumn{1}{c|}{\textbf{specificity}} & \textbf{accuracy} \\ \hline
\textbf{SVM}                                                                       & \multicolumn{1}{c|}{0.866}                & \multicolumn{1}{c|}{0.738}                & 0.800             & \multicolumn{1}{c|}{0.903}                & \multicolumn{1}{c|}{0.667}                & 0.787             \\ \hline
\textbf{\begin{tabular}[c]{@{}c@{}}AdaBoost-SAMME w/\\ Random Forest\end{tabular}} & \multicolumn{1}{c|}{0.938}                & \multicolumn{1}{c|}{0.874}                & 0.905             & \multicolumn{1}{c|}{0.903}                & \multicolumn{1}{c|}{0.833}                & 0.869             \\ \hline
\textbf{XGBoost}                                                                   & \multicolumn{1}{c|}{0.897}                & \multicolumn{1}{c|}{0.864}                & 0.880             & \multicolumn{1}{c|}{0.935}                & \multicolumn{1}{c|}{0.800}                & 0.869             \\ \hline
\end{tabular}
\end{table*}

As seen above, using a weekly aggregation time window leads to better model performance when compared with a monthly aggregation time window. However, since both datasets were collected for one month each, when using the monthly window we obtain one classification per student. On the other hand, when using the weekly time window, each student has several classifications, that is one per each week of data. However, when considering the real-world context, each student only has one final classification. As such, in this section we propose a method to not only output a final classification per student, but, additionally, to improve the previous models' performance. Below we present the tests for this method using the weekly aggregation time window, the joined dataset, and the three best performing models from last section.   

Models to merge classification, have been widely covered in past literature\cite{jahrer2010combining}. However, this strategy is mostly used to combine the output of multiple models into one final classification at decision level. This strategy has several possible implementations, namely by computing the median value of the classifications, the majority vote, the product of probabilities, the average of probabilities, the minimum probability, or the maximum probability. Our approach differs from these strategies since it only uses one model to perform the classification of instances. Specifically, we propose an approach that combines all outputs of the model for the same student, to output a final classification per student. 

The pipeline of the proposed mechanism can be seen in Figure \ref{fig:pipeline_median}. In the first part of the pipeline, each sample is processed as an independent entry for the model. After obtaining a prediction for each sample, the samples are grouped by student and processed by the median voting algorithm. This algorithm then outputs a final classification, that is the same classification for every sample of the same student. The only requirement for this mechanism to work, is that each sample's student identification must be carried on to the final classification step. Additionally, this mechanism will work with any number of samples per student, equal to or greater than 1. 

The median voting algorithm can be seen in Algorithm \ref{alg:median_vote}. The algorithm requires as inputs the list of predictions of the model and the student identification for each prediction in the same order and with the same size.  The algorithm then iterates over the list of predictions and replaces the value of each prediction with the median for the subset of predictions of that prediction's student. When the student has the same number of classifications for each class (e.g., 2 predictions of \textit{above median} and 2 predictions of \textit{below median}), the \textit{above median} class is taken, since we are employing the condition of greater or equal to 0.5. Furthermore, the final predictions should be allocated to a new array, so the median value does not change with every iteration. The algorithm then outputs the final predictions for each sample.

Although simple, this method can improve the previously discussed models' performance. This is due to the fact that even \textit{good} students (that would be classified as \textit{above median}) can have \textit{bad} weeks. A similar thing can happen in the opposite case, that is, a student that performed \textit{below median} can have an \textit{above median} week. However, that \textit{good} week might not be enough to account for the other weeks that negatively affected the performance.

We tested this mechanism for the three models that presented better performance in the previous section, namely \gls{SVM}, AdaBoost and XGBoost. The results of these models with the voting algorithm can be seen in Table \ref{tab:performance_pipeline}. We can see that this pipeline effectively increases the performance of all models. In the specific case of XGBoost and AdaBoost, the accuracy increase is in excess of 10\%. We can also see that the AdaBoost model is the one that offers the best performance, with an accuracy of 90.5\%, followed by XGBoost with 88\% accuracy. We can also see that the \gls{SVM} model was the only one for which one of the evaluated metrics decreased. Specifically, \gls{SVM}'s its specificity decreased by 1\%. Additionally, we can see that all models favour the \textit{above median} class (i.e., the sensitivity is higher than the specificity). This is due to the use of the condition of greater or equal to 0.5, in Algorithm \ref{alg:median_vote}, in the case there is a tie between the number of predictions for both classes for one student. However, depending on the case-study (e.g., a higher False Positive Rate is acceptable), this condition can be changed in order to favour the false class.

Using this pipeline, we get the same classification for every instance of data of the same student. As such, this effectively gives us one classification per student. In Table \ref{tab:performance_pipeline}, we also present the metrics of each model when computed by student. We can see that, once again, AdaBoost and XGBoost offer a better performance, being able to correctly classify 53 of the total 61 students. When comparing these results with the classification using a monthly time window, we can see that the use of this pipeline leads to a big increase in performance. As such, we believe that this approach can be used to create better models to automatically access students' performance. 

Additionally, since this pipeline uses several instances of data from each student, to reach a final classification per student it can in fact improve over time. For that, the model only needs to keep a history of instances of each student. That is, envisioning a model that retrieves data and outputs a classification per student every week, as time progresses the model will have a larger number of instances per student. That means that every week the classification given to the student can be more accurate.

\section{Conclusion}
\label{sec:conclusion}

In this paper, we presented an approach to predict students’ academic performance, based both on automatically collected data using the \gls{ISABELA} \gls{HITLCPS} system, and on a questionnaire-based approach to collect emotional state data. The objective of the prediction was to classify the students as performing above median or below median.

Several prediction models were developed and studied, using datasets collected during two distinct one-month trials, involving 61 students and two different school subjects, executed in non-Covid (2018 dataset) and Covid (2021 dataset) periods, in Ecuador’s Escuela Politecnica Nacional. Extensive results point to the fact that SVM, AdaBoost, and XGBoost, are the best performing models for this classification problem. Furthermore, we evaluated the use of two different data aggregation time windows to predict student performance. The results show that the use of a weekly time window can have several advantages over the use of a monthly time window.

Although the datasets were acquired under significantly different situations, in addition to performing tests with the individual datasets we also performed tests using the joined dataset. Results show that the joined dataset create a more generalize model that is able to predict instances from both periods of time with 79\% accuracy.

As a final study, we propose a pipeline that uses a median voting algorithm at decision level to further increase the model’s performance. The results show that this approach is able to increase the performance of selected models by more than 10\%. 

The obtained results are promising in what concerns the creation of a model that is able to predict students’ academic performance. Results also show that the created models are able to deal with changes in the context of learning (i.e., pandemic versus non-pandemic context). However, due to small size of the dataset we cannot draw general conclusions for a larger population. More data must be collected, for different school subjects and with more students, in order to evaluate the validity of these results. In our future work, we intend to perform more data acquisition trials, and further evaluate our proposed pipeline. Furthermore, the proposed classification method, based on the median value, offers a mechanism that should work for different regions of the world despite different classification methodologies. However, more tests should be done with datasets collected in different parts of the world to further support the results presented in this paper.


\ifCLASSOPTIONcompsoc
  \section*{Acknowledgments}
\else
  \section*{Acknowledgment}
\fi

This work is funded by the FCT - Foundation for Science and Technology, I.P./MCTES through national funds (PIDDAC), within the scope of CISUC R\&D Unit - UIDB/00326/2020 or project code UIDP/00326/2020. José Marcelo Fernandes wishes to acknowledge the Portuguese funding institution FCT - Foundation for Science and Technology for supporting his research under the Ph.D. grants SFRH/BD/147371/2019.



%


%

\bibliographystyle{ieeetr}
\bibliography{bibliography}

\begin{thebibliography}{10}

\bibitem{IoTconn92:online}
``• iot connected devices worldwide 2019-2030 | statista.''
  \url{https://www.statista.com/statistics/1183457/iot-connected-devices-worldwide/}.
\newblock (Accessed on 05/19/2022).

\bibitem{nunes2015survey}
D.~S. Nunes, P.~Zhang, and J.~S. Silva, ``A survey on human-in-the-loop
  applications towards an internet of all,'' {\em IEEE Communications Surveys
  \& Tutorials}, vol.~17, no.~2, pp.~944--965, 2015.

\bibitem{hijazi2006factors}
S.~T. Hijazi and S.~Naqvi, ``Factors affecting students' performance.,'' {\em
  Bangladesh e-journal of Sociology}, vol.~3, no.~1, 2006.

\bibitem{strelan2020flipped}
P.~Strelan, A.~Osborn, and E.~Palmer, ``The flipped classroom: A meta-analysis
  of effects on student performance across disciplines and education levels,''
  {\em Educational Research Review}, vol.~30, p.~100314, 2020.

\bibitem{haleem2020effects}
A.~Haleem, M.~Javaid, and R.~Vaishya, ``Effects of covid-19 pandemic in daily
  life,'' {\em Current medicine research and practice}, vol.~10, no.~2, p.~78,
  2020.

\bibitem{buzzi2020psycho}
C.~Buzzi, M.~Tucci, R.~Ciprandi, I.~Brambilla, S.~Caimmi, G.~Ciprandi, and
  G.~L. Marseglia, ``The psycho-social effects of covid-19 on italian
  adolescents’ attitudes and behaviors,'' {\em Italian journal of
  pediatrics}, vol.~46, no.~1, pp.~1--7, 2020.

\bibitem{afonso2020impact}
P.~Afonso, ``The impact of the covid-19 pandemic on mental health,'' {\em Acta
  medica portuguesa}, vol.~33, no.~5, pp.~356--357, 2020.

\bibitem{fernandes2020isabela}
J.~Fernandes, D.~Raposo, N.~Armando, S.~Sinche, J.~S. Silva, A.~Rodrigues,
  V.~Pereira, H.~G. Oliveira, L.~Macedo, and F.~Boavida, ``Isabela--a
  socially-aware human-in-the-loop advisor system,'' {\em Online Social
  Networks and Media}, vol.~16, p.~100060, 2020.

\bibitem{wang2014studentlife}
R.~Wang, F.~Chen, Z.~Chen, T.~Li, G.~Harari, S.~Tignor, X.~Zhou, D.~Ben-Zeev,
  and A.~T. Campbell, ``Studentlife: assessing mental health, academic
  performance and behavioral trends of college students using smartphones,'' in
  {\em Proceedings of the 2014 ACM international joint conference on pervasive
  and ubiquitous computing}, pp.~3--14, 2014.

\bibitem{shiffman2008ecological}
S.~Shiffman, A.~A. Stone, and M.~R. Hufford, ``Ecological momentary
  assessment,'' {\em Annu. Rev. Clin. Psychol.}, vol.~4, pp.~1--32, 2008.

\bibitem{wang2015smartgpa}
R.~Wang, G.~Harari, P.~Hao, X.~Zhou, and A.~T. Campbell, ``Smartgpa: how
  smartphones can assess and predict academic performance of college
  students,'' in {\em Proceedings of the 2015 ACM international joint
  conference on pervasive and ubiquitous computing}, pp.~295--306, 2015.

\bibitem{tibshirani1996regression}
R.~Tibshirani, ``Regression shrinkage and selection via the lasso,'' {\em
  Journal of the Royal Statistical Society: Series B (Methodological)},
  vol.~58, no.~1, pp.~267--288, 1996.

\bibitem{harari2017patterns}
G.~M. Harari, S.~D. Gosling, R.~Wang, F.~Chen, Z.~Chen, and A.~T. Campbell,
  ``Patterns of behavior change in students over an academic term: A
  preliminary study of activity and sociability behaviors using smartphone
  sensing methods,'' {\em Computers in Human Behavior}, vol.~67, pp.~129--138,
  2017.

\bibitem{osmanbegovic2012data}
E.~Osmanbegovic and M.~Suljic, ``Data mining approach for predicting student
  performance,'' {\em Economic Review: Journal of Economics and Business},
  vol.~10, no.~1, pp.~3--12, 2012.

\bibitem{nepal2022covid}
S.~Nepal, W.~Wang, V.~Vojdanovski, J.~F. Huckins, A.~daSilva, M.~Meyer, and
  A.~Campbell, ``Covid student study: A year in the life of college students
  during the covid-19 pandemic through the lens of mobile phone sensing,'' in
  {\em CHI Conference on Human Factors in Computing Systems}, pp.~1--19, 2022.

\bibitem{sandoval2020early}
I.~Sandoval-Palis, D.~Naranjo, J.~Vidal, and R.~Gilar-Corbi, ``Early dropout
  prediction model: A case study of university leveling course students,'' {\em
  Sustainability}, vol.~12, no.~22, p.~9314, 2020.

\bibitem{activityRecognition}
``Activity recognition api  |  google developers.''
  \url{https://developers.google.com/location-context/activity-recognition}.
\newblock (Accessed on 06/02/2022).

\bibitem{Bynion2020}
T.-M. Bynion and M.~T. Feldner, {\em Self-Assessment Manikin}, pp.~4654--4656.
\newblock Cham: Springer International Publishing, 2020.

\bibitem{yazdani2013multimedia}
A.~Yazdani, E.~Skodras, N.~Fakotakis, and T.~Ebrahimi, ``Multimedia content
  analysis for emotional characterization of music video clips,'' {\em EURASIP
  Journal on Image and Video Processing}, vol.~2013, no.~1, pp.~1--10, 2013.

\bibitem{taras2005sleep}
H.~Taras and W.~Potts-Datema, ``Sleep and student performance at school,'' {\em
  Journal of school health}, vol.~75, no.~7, pp.~248--254, 2005.

\bibitem{quinlan1996learning}
J.~R. Quinlan, ``Learning decision tree classifiers,'' {\em ACM Computing
  Surveys (CSUR)}, vol.~28, no.~1, pp.~71--72, 1996.

\bibitem{breiman2001random}
L.~Breiman, ``Random forests,'' {\em Machine learning}, vol.~45, no.~1,
  pp.~5--32, 2001.

\bibitem{mammone2009support}
A.~Mammone, M.~Turchi, and N.~Cristianini, ``Support vector machines,'' {\em
  Wiley Interdisciplinary Reviews: Computational Statistics}, vol.~1, no.~3,
  pp.~283--289, 2009.

\bibitem{rish2001empirical}
I.~Rish {\em et~al.}, ``An empirical study of the naive bayes classifier,'' in
  {\em IJCAI 2001 workshop on empirical methods in artificial intelligence},
  vol.~3, pp.~41--46, 2001.

\bibitem{cover1967nearest}
T.~Cover and P.~Hart, ``Nearest neighbor pattern classification,'' {\em IEEE
  transactions on information theory}, vol.~13, no.~1, pp.~21--27, 1967.

\bibitem{hastie2009multi}
T.~Hastie, S.~Rosset, J.~Zhu, and H.~Zou, ``Multi-class adaboost,'' {\em
  Statistics and its Interface}, vol.~2, no.~3, pp.~349--360, 2009.

\bibitem{chen2016xgboost}
T.~Chen and C.~Guestrin, ``Xgboost: A scalable tree boosting system,'' in {\em
  Proceedings of the 22nd acm sigkdd international conference on knowledge
  discovery and data mining}, pp.~785--794, 2016.

\bibitem{jahrer2010combining}
M.~Jahrer, A.~T{\"o}scher, and R.~Legenstein, ``Combining predictions for
  accurate recommender systems,'' in {\em Proceedings of the 16th ACM SIGKDD
  international conference on Knowledge discovery and data mining},
  pp.~693--702, 2010.

\end{thebibliography}
\newpage


\section{appendix}

\begin{table}[h]
\centering
\renewcommand{\arraystretch}{1.3}
\caption{Grid search parameters for the K-Near-Neighbors model, with selected configurations highlighted for each dataset.}
\label{tab:grid_knn}
\begin{tabular*}{0.45\textwidth}{@{\extracolsep{\fill}}|c|c|c|r|r|}
\hline
                         & \textit{\textbf{algorithm}}                                                   & \textit{\textbf{weights}}                                  & \textit{\textbf{p}}                           & \textit{\textbf{n\_neighbors}}                                    \\ \hline
\textit{\textbf{2018}}   & \begin{tabular}[c]{@{}c@{}}\textbf{\textit{\underline{auto}}}\\ ball\_tree \\ kd\_tree\\ brute\end{tabular} & \begin{tabular}[c]{@{}c@{}}\textbf{\textit{\underline{uniform}}}\\ distance\end{tabular} & \begin{tabular}[c]{@{}r@{}}\textbf{\textit{\underline{1}}}\\ 2\end{tabular} & \begin{tabular}[c]{@{}r@{}}1\\\textbf{\textit{\underline{2}}}\\ 3\\ 4\\ 5\\ 6\\ 7\end{tabular} \\ \hline
\textit{\textbf{2021}}   & \begin{tabular}[c]{@{}c@{}}\textbf{\textit{\underline{auto}}}\\ ball\_tree \\ kd\_tree\\ brute\end{tabular} & \begin{tabular}[c]{@{}c@{}}uniform\\ \textbf{\textit{\underline{distance}}}\end{tabular} & \begin{tabular}[c]{@{}r@{}}1\\ \textbf{\textit{\underline{2}}}\end{tabular} & \begin{tabular}[c]{@{}r@{}}1\\ 2\\ 3\\ \textbf{\textit{\underline{4}}}\\ 5\\ 6\\ 7\end{tabular} \\ \hline
\textit{\textbf{Joined}} & \begin{tabular}[c]{@{}c@{}}\textbf{\textit{\underline{auto}}}\\ ball\_tree \\ kd\_tree\\ brute\end{tabular} & \begin{tabular}[c]{@{}c@{}}uniform\\ \textbf{\textit{\underline{distance}}}\end{tabular} & \begin{tabular}[c]{@{}r@{}}\textbf{\textit{\underline{1}}}\\ 2\end{tabular} & \begin{tabular}[c]{@{}r@{}}1\\ 2\\ 3\\ 4\\ 5\\ \textbf{\textit{\underline{6}}}\\ 7\end{tabular} \\ \hline
\end{tabular*}
\end{table}

\begin{table}[h]
\centering
\renewcommand{\arraystretch}{1.3}
\caption{Grid search parameters for the XGBoost model, with selected configurations highlighted for each dataset.}
\label{tab:grid_xgb}
\begin{tabular*}{0.45\textwidth}{@{\extracolsep{\fill}}|c|c|r|r|}
\hline
                         & \textit{\textbf{booster}}                                      & \multicolumn{1}{c|}{\textit{\textbf{n\_estimators}}}             & \multicolumn{1}{c|}{\textit{\textbf{learning\_rate}}}                   \\ \hline
\textit{\textbf{2018}}   & \begin{tabular}[c]{@{}c@{}}\textbf{\textit{\underline{gbtree}}}\\ linear\\ dart\end{tabular} & \begin{tabular}[c]{@{}r@{}}\textbf{\textit{\underline{20}}}\\ 50\\ 75\\ 100\\ 200\end{tabular} & \begin{tabular}[c]{@{}r@{}}0.1\\ 0.5\\ \textbf{\textit{\underline{0.6}}}\\ 0.7\\ 0.8\\ 1\end{tabular} \\ \hline
\textit{\textbf{2021}}   & \begin{tabular}[c]{@{}c@{}}\textbf{\textit{\underline{gbtree}}}\\ linear\\ dart\end{tabular} & \begin{tabular}[c]{@{}r@{}}20\\ \textbf{\textit{\underline{50}}}\\ 75\\ 100\\ 200\end{tabular} & \begin{tabular}[c]{@{}r@{}}0.1\\ 0.5\\ \textbf{\textit{\underline{0.6}}}\\ 0.7\\ 0.8\\ 1\end{tabular} \\ \hline
\textit{\textbf{Joined}} & \begin{tabular}[c]{@{}c@{}}\textbf{\textit{\underline{gbtree}}}\\ linear\\ dart\end{tabular} & \begin{tabular}[c]{@{}r@{}}20\\ 50\\ 75\\ 100\\ \textbf{\textit{\underline{200}}}\end{tabular} & \begin{tabular}[c]{@{}r@{}}0.1\\ 0.5\\ 0.6\\ 0.7\\ \textbf{\textit{\underline{0.8}}}\\ 1\end{tabular} \\ \hline
\end{tabular*}
\end{table}

\begin{table}[h]
\centering
\renewcommand{\arraystretch}{1.3}
\caption{Grid search parameters for the SVM model, with selected components highlighted for each dataset.}
\label{tab:grid_svm}
\begin{tabular*}{0.45\textwidth}{@{\extracolsep{\fill}}|c|c|r|c|}
\hline
                         & \textit{\textbf{Kernel}}                                              & \multicolumn{1}{c|}{\textit{\textbf{C}}}                    & \textit{\textbf{Gamma}}                              \\ \hline
\textit{\textbf{2018}}   & \begin{tabular}[c]{@{}c@{}}poly\\ linear\\ sigmoid\\ \underline{\textit{\textbf{rbf}}}\end{tabular} & \begin{tabular}[c]{@{}r@{}}\underline{\textbf{\textit{1}}}\\ 10\\ 100\\ 1000\end{tabular} & \begin{tabular}[c]{@{}c@{}}\underline{\textbf{\textit{auto}}}\\ scale\end{tabular} \\ \hline
\textit{\textbf{2021}}   & \begin{tabular}[c]{@{}c@{}}poly\\ linear\\ sigmoid\\ \underline{\textit{\textbf{rbf}}}\end{tabular} & \begin{tabular}[c]{@{}r@{}}1\\ \underline{\textbf{\textit{10}}}\\ 100\\ 1000\end{tabular} & \begin{tabular}[c]{@{}c@{}}\underline{\textbf{\textit{auto}}}\\ scale\end{tabular} \\ \hline
\textit{\textbf{Joined}} & \begin{tabular}[c]{@{}c@{}}poly\\ linear\\ sigmoid\\ \underline{\textit{\textbf{rbf}}}\end{tabular} & \begin{tabular}[c]{@{}r@{}}\underline{\textbf{\textit{1}}}\\ 10\\ 100\\ 1000\end{tabular} & \begin{tabular}[c]{@{}c@{}}\underline{\textbf{\textit{auto}}}\\ scale\end{tabular} \\ \hline
\end{tabular*}
\end{table}

\begin{table}[h]
\caption{Grid search parameters for the Naibe Bayes model, with selected components highlighted for each dataset.}
\centering
\renewcommand{\arraystretch}{1.4}
\label{tab:grid_nb}
\begin{tabular}{|c|c|}
\hline
                         & \textit{\textbf{var\_smothing}}                                                                           \\ \hline
\textit{\textbf{2018}}   & \begin{tabular}[c]{@{}c@{}}1e-5\\ 1e-6\\ 1e-7\\ 1e-8\\ 1e-9\\ 1e-10\\ \textbf{\textit{\underline{1e-11}}} \\ 1e-12\\ 1e-15\end{tabular} \\ \hline
\textit{\textbf{2021}}   & \begin{tabular}[c]{@{}c@{}}1e-5\\ 1e-6\\ 1e-7\\ 1e-8\\ 1e-9\\ 1e-10\\ \textbf{\textit{\underline{1e-11}}} \\ 1e-12\\ 1e-15\end{tabular} \\ \hline
\textit{\textbf{Joined}} & \begin{tabular}[c]{@{}c@{}}1e-5\\ 1e-6\\ 1e-7\\ 1e-8\\ 1e-9\\ 1e-10\\ \textbf{\textit{\underline{1e-11}}} \\ 1e-12\\ 1e-15\end{tabular} \\ \hline
\end{tabular}
\end{table}

\begin{table*}[h!]
\centering
\renewcommand{\arraystretch}{1.3}
\caption{Grid search parameters for the Random Forest model, with selected configurations highlighted for each dataset.}
\label{tab:grid_rf}
\begin{tabular}{|c|c|r|c|r|r|r|}
\hline
                         & \textit{\textbf{bootstrap}}                          & \multicolumn{1}{c|}{\textit{\textbf{n\_estimators}}}        & \textit{\textbf{max\_features}}                             & \multicolumn{1}{c|}{\textit{\textbf{max\_depth}}}                                                   & \multicolumn{1}{c|}{\textit{\textbf{min\_samples\_leaf}}} & \multicolumn{1}{c|}{\textit{\textbf{min\_samples\_split}}} \\ \hline
\textit{\textbf{2018}}   & \begin{tabular}[c]{@{}c@{}}\textbf{\textit{\underline{True}}}\\ False\end{tabular} & \begin{tabular}[c]{@{}r@{}}10\\ \textbf{\textit{\underline{50}}}\\ 100\\ 200\end{tabular} & \begin{tabular}[c]{@{}c@{}}log2\\ sqrt \\ \textbf{\textit{\underline{None}}}\end{tabular} & \begin{tabular}[c]{@{}r@{}}10\\ 20\\ 30\\ 40\\ 50\\ 60\\ 70\\  80\\  90\\  100\\  \textbf{\textit{\underline{None}}}\end{tabular} & \begin{tabular}[c]{@{}r@{}}1\\ \textbf{\textit{\underline{2}}}\\ 4\end{tabular}         & \begin{tabular}[c]{@{}r@{}}2\\ 5\\ \textbf{\textit{\underline{10}}}\end{tabular}         \\ \hline
\textit{\textbf{2021}}   & \begin{tabular}[c]{@{}c@{}}\textbf{\textit{\underline{True}}}\\ False\end{tabular} & \begin{tabular}[c]{@{}r@{}}10\\ 50\\ \textbf{\textit{\underline{100}}}\\ 200\end{tabular} & \begin{tabular}[c]{@{}c@{}}log2\\ \textbf{\textit{\underline{sqrt}}} \\ None\end{tabular} & \begin{tabular}[c]{@{}r@{}}10\\ 20\\ 30\\ 40\\ 50\\ \textbf{\textit{\underline{60}}}\\ 70\\  80\\  90\\  100\\  None\end{tabular} & \begin{tabular}[c]{@{}r@{}}1\\ \textbf{\textit{\underline{2}}}\\ 4\end{tabular}         & \begin{tabular}[c]{@{}r@{}}2\\ 5\\ \textbf{\textit{\underline{10}}}\end{tabular}         \\ \hline
\textit{\textbf{Joined}} & \begin{tabular}[c]{@{}c@{}}True\\ \textbf{\textit{\underline{False}}}\end{tabular} & \begin{tabular}[c]{@{}r@{}}10\\ 50\\ \textbf{\textit{\underline{100}}}\\ 200\end{tabular} & \begin{tabular}[c]{@{}c@{}}log2\\ \textbf{\textit{\underline{sqrt}}} \\ None\end{tabular} & \begin{tabular}[c]{@{}r@{}}10\\ 20\\ 30\\ 40\\ 50\\ 60\\ 70\\  80\\  \textbf{\textit{\underline{90}}}\\  100\\  None\end{tabular} & \begin{tabular}[c]{@{}r@{}}\textbf{\textit{\underline{1}}}\\ 2\\ 4\end{tabular}         & \begin{tabular}[c]{@{}r@{}}2\\ 5\\ \textbf{\textit{\underline{10}}}\end{tabular}         \\ \hline
\end{tabular}
\end{table*}

\begin{table*}[h!]
\centering
\renewcommand{\arraystretch}{1.3}
\caption{Grid search parameters for the Decision Tree model, with selected configurations highlighted for each dataset.}
\label{tab:grid_dt}
\begin{tabular}{|c|c|r|r|r|r|}
\hline
                         & \textit{\textbf{criterion}}                            & \multicolumn{1}{c|}{\textit{\textbf{max\_depth}}}                                                           & \multicolumn{1}{c|}{\textit{\textbf{max\_features}}}                                                       & \multicolumn{1}{l|}{\textit{\textbf{min\_samples\_leaf}}} & \multicolumn{1}{l|}{\textit{\textbf{min\_samples\_split}}} \\ \hline
\textit{\textbf{2018}}   & \begin{tabular}[c]{@{}c@{}}\textbf{\textit{\underline{gini}}}\\ entropy\end{tabular} & \begin{tabular}[c]{@{}r@{}}5\\ 8\\ 10\\ 20\\ 30\\ 40\\ 50\\ 60\\ 70\\  80\\ \textbf{\textit{\underline{90}}}\\  100\\  None\end{tabular} & \begin{tabular}[c]{@{}r@{}}5\\  6\\  7\\  8\\  9\\  10\\  11\\  12\\  13\\ \textbf{\textit{\underline{sqrt}}}\\ log2\\ None\end{tabular} & \begin{tabular}[c]{@{}r@{}}\textbf{\textit{\underline{1}}}\\ 2\\ 4\end{tabular}         & \begin{tabular}[c]{@{}r@{}}2\\ \textbf{\textit{\underline{5}}}\\ 10\end{tabular}         \\ \hline
\textit{\textbf{2021}}   & \begin{tabular}[c]{@{}c@{}}\textbf{\textit{\underline{gini}}}\\ entropy\end{tabular} & \begin{tabular}[c]{@{}r@{}}\textbf{\textit{\underline{5}}}\\ 8\\ 10\\ 20\\ 30\\ 40\\ 50\\ 60\\ 70\\  80\\  90\\  100\\  None\end{tabular} & \begin{tabular}[c]{@{}r@{}}5\\  6\\  7\\  8\\  9\\  10\\  11\\  12\\  13\\ \textbf{\textit{\underline{sqrt}}}\\ log2\\ None\end{tabular} & \begin{tabular}[c]{@{}r@{}}\textbf{\textit{\underline{1}}}\\ 2\\ 4\end{tabular}         & \begin{tabular}[c]{@{}r@{}}2\\\textbf{\textit{\underline{5}}}\\ 10\end{tabular}         \\ \hline
\textit{\textbf{Joined}} & \begin{tabular}[c]{@{}c@{}}\textbf{\textit{\underline{gini}}}\\ entropy\end{tabular} & \begin{tabular}[c]{@{}r@{}}5\\ 8\\ 10\\ 20\\ 30\\ 40\\ 50\\ 60\\ 70\\  80\\  90\\  100\\  \textbf{\textit{\underline{None}}}\end{tabular} & \begin{tabular}[c]{@{}r@{}}5\\  6\\  7\\  8\\  \textbf{\textit{\underline{9}}}\\  10\\  11\\  12\\  13\\ sqrt\\ log2\\ None\end{tabular} & \begin{tabular}[c]{@{}r@{}}1\\ 2\\ \textbf{\textit{\underline{4}}}\end{tabular}         & \begin{tabular}[c]{@{}r@{}}2\\ 5\\ \textbf{\textit{\underline{10}}}\end{tabular}         \\ \hline
\end{tabular}
\end{table*}


\end{document}